\documentclass{article}
\usepackage[T1]{fontenc}
\usepackage[utf8]{inputenc}
\usepackage[italian,english]{babel}
\usepackage{amsmath}
\usepackage{amsthm}
\usepackage{amssymb}
\usepackage{comment}
\usepackage{eucal}
\usepackage{graphicx}
\usepackage{enumerate}
\usepackage{pictex, dcpic}
\usepackage[babel]{csquotes}
\usepackage[style=numeric,sorting=none,sortcites=true]{biblatex}

\usepackage{color}

\def\al{\alpha}
\def\be{\beta}
\def\de{\delta}
\def\ga{\gamma}

\def\ep{\epsilon}

\def\te{\theta}
\def\la{\lambda}
\def\ze{\zeta}
\def\om{\omega}
\def\si{\sigma}
\def\vp{\varphi}

\def\Ga{\Gamma}

 \def\R{{{\mathbb R}}}

 \def\R{{{\mathbb R}}}

\def\Met{{\hbox{Met}}}

\def\Aut{{\hbox{Aut}}}

\def\GL{{\hbox{GL}}}

\def\Diff{{\hbox{Diff}}}

\def\ip{\hbox to4pt{\leaders\hrule height0.3pt\hfill}\vbox to8pt{\leaders\vrule width0.3pt\vfill}\kern 2pt}
\def\del{\partial}
\def\na{\nabla}

\def\Lie{\pounds}

\def\arr{\rightarrow}

\def\ffrac[#1/#2]{\hbox{$\frac{#1}{#2}$}}
\def\Frac[#1/#2]{\frac{#1}{#2}}
\def\({\left(}
\def\){\right)}
\def\[{\left[}
\def\]{\right]}
\def\^#1{{}^{#1}_{\>\cdot}}
\def\_#1{{}_{#1}^{\>\cdot}}
\def\<{\kern -1pt}

\def\beq{\begin{equation}}
\def\eeq{\end{equation}}
{\left\lbrace\begin{array}{@{}l@{}}}%
{\end{array}\right.}
\def\Noe{\mathcal{E}}
\def\UU{\mathcal{U}}

\begin{document}

\title{Gauge Natural Formulation of \\ Conformal Gravity}
\author{M.Campigotto$^{a,b}$, L.Fatibene$^{c,b}$
\\
{\small $^a$ Dipartimento di Fisica, University of Torino, Via P. Giuria 1, 10125, Torino, Italy}\\
{\small $^b$ Istituto Nazionale di Fisica Nucleare (INFN), Via P. Giuria 1, 10125, Torino, Italy}\\
{\small $^c$ Dipartimento di Matematica, University of Torino, Via C. Alberto 10, 10123, Torino, Italy}\\
}

\maketitle

\begin{abstract}
We consider conformal gravity as a gauge natural theory. We study its conservation laws and  superpotentials. We also consider the Mannheim and Kazanas spherically symmetric vacuum solution and discuss  conserved quantities associated to conformal and diffeomorphism symmetries.
\end{abstract}

\section{Introduction}
General Relativity is a self-consistent covariant  theory for gravity which is able to successfully describe gravity at many scales. 
Its predictions agree with observations at the Solar System and astrophysical scales.
However, at galactic and cosmological scales, one needs to introduce a huge amount of dark sources in the form of dark matter and dark energy
in order to  model phenomenological aspects such as galaxy rotation curves or properly describe structures formation. 
On the other hand what dark sources are at fundamental level remains a mystery.
While looking for fundamental models for dark sources is an option, one can consider desirable alternative theories of gravity which account for dark sources effectively as pure gravitational effects.

Philip Mannheim proposed a conformally invariant theory of gravity based on the conformal Weyl tensor \cite{Mannheim:1989}. 
This theory is worth being considered among the class of modified or extended gravitational models.
Besides being a candidate for physical modeling of gravitational phenomena at different scales it also provides a test for a more general understanding of the relation between covariant theories and observations.
We shall show that conformal gravity  can be included in the more general framework of gauge natural theories adding conformal invariance right into the game at kinematical level. In this way the theory is formally very similar to a gauge theory and the geometrical meaning of its fields is rendered explicitly and suitably encoded into a principal bundle over the spacetime manifold; see \cite{Jackiv}  for a more detailed discussion about the physical meaning of these conformal transformations in similar though different kinds of theories.

Depending on the context, two related but distinct physical transformations
are denoted as {\it conformal} in the literature. While conformal transformations are often considered (as in conformal field theories) as particular coordinate transformations which leave the metric structure unchanged modulo a (positive) factor, they are sometimes also introduced (especially in conformal gravity) as transformations which do not affect the spacetime point while they change the metric field by a pointwise (positive) factor. According to the first viewpoint conformal transformations are a special class of diffeomorphisms.
In the second viewpoint conformal transformations are gauge transformations acting on fields alone (i.e.~they are vertical transformations on the configuration bundle).  

The differences between these two viewpoints may be considered trivial though they have a certain amount of consequences which are worth addressing.
A trivial difference is that conformal transformations on spacetime (i.e.~the first viewpoint) form a group which is bigger than isometries and smaller than diffeomorphisms. While it can be meaningful to try and extend a special relativistic theory to be covariant which respect to the bigger group of these conformal transformations (as essentially one does in conformal field theories) it makes relatively no sense to consider a conformal version of a generally covariant theory (as one does in conformal gravity). A generally covariant theory is already covariant with respect to all diffeomorphisms (including the subgroup of conformal spacetime transformations).
In other words, in conformal gravity the only allowed viewpoint is to regard conformal transformations as gauge transformations. 
This of course produces ambiguous notations which need to be dealt with care.
It is our opinion that the gauge natural approach provides a good foundation of both viewpoints and it allows to avoid notational ambiguities.
However, hereafter we shall not discuss this issue in detail. We shall only present the gauge natural formulation of conformal gravity and discuss conservation laws, leaving the foundational issues for a future more general investigation.

As we shall see, Weyl tensor comes in quite naturally as a consequence of symmetry requirements on dynamics and a canonical treatment of conservation laws is a free token from gauge natural framework (see \cite{Opava}, \cite{GaugeNatural}, \cite{FatibeneLibro}). 
Conserved currents for gauge natural theories are exact differential forms (on-shell), which do admit a superpotential. Thus, it can be developed a canonical way of finding conserved quantities.
Therefore, after having set up a well-founded geometrical framework, one has a useful tool for analyzing physical phenomena such as the gravitational lensing.
%Therefore, once one has set up a well-founded geometrical framework, that could be a useful tool for analyzing physical phenomena such as the gravitational lensing.

Especially in conformal gravity the issue of conserved quantities would be particularly important to be understood generally for applications for example to gravitational lensing (and this paper is meant to be in preparation for such an analysis).
In fact solutions in conformal gravity are particularly poorly understood. First of all  being the theory conformally invariant any weak field approximation will still be conformally invariant. For example one obtains a good theory of motion of test light rays (which is in fact conformally invariant)
while the Newtonian limit of test particles would depend on the details of some gauge fixing. Since the masses as defined in astrophysics are definitely not conformally invariant (the third Kepler law is definitely not conformally invariant) these notions cannot simply be obtained by Newtonian limit (or at least not {\it only} in terms of Newtonian limit but they would  depend on the details of some gauge fixing).
The same attention needs to be paid if one defines masses in terms of conservation laws (as we shall see the superpotentials are in fact conformally invariant as their integrals will be).

On the other hand in solutions which are not asymptotically flat (as they are in conformal gravity but also in many cases of GR models)
a great effort has to be spent in order to define deflection angles so that they make geometrical sense. One way would be to compute the angle between received light rays with and without the gravitational lens. 
This procedure in particular needs a complete control on what it means to {\it switch off the lens}, i.e.~which parameters appearing in the solution are related to the localized source and which are related to the asymptotic behavior.
This complete control is still to be obtained, in conformal gravity as in other models, and should be based on a careful analysis of conservation laws which is what we start to do hereafter.

\section{Gauge Natural Framework}

In general, a gauge natural theory is defined to be a field theory in which fields are sections of a gauge natural bundle $C$ associated to a principal bundle $P$, in which the dynamics is covariant with respect to gauge transformations defined as automorphisms of $P$. Gauge transformations canonically act
on the associated configuration bundle $C$. Moreover, all sections of $C$, namely all fields, are required to be dynamical. We refer to \cite{FatibeneLibro} for general notation and framework.
This framework has proven to be suitable to discuss gauge theories in their generally relativistic formulations, as well as couplings between spinor fields and gravity; see \cite{Spinors}. 

Hereafter we shall specialize this general framework to the case of conformal theory of gravitation. By showing that conformal gravity fits in the framework of gauge natural theories one gets for free a canonical treatment of conservation laws (as well as a strong control on global properties of fields and their observability).

Let us start by considering a (connected, paracompact) manifold $M$ of dimension $\dim(M)=m$ which allows global metrics of signature $\eta=(r,s)$ (of course with $m=r+s$). 
The manifold $M$ is a model for spacetime, though {\it before} fixing a specific metric on it.

Let $P$ be a principal bundle on $M$ with the (abelian) group $G=(\R,+)$. Local fibered coordinates on $P$ are $(x^\mu, l)$ and two such coordinate systems are related by transition functions in this form
\begin{equation}
\begin{cases}
x'^\mu= x'^\mu(x)\\
l'= \al(x) +l
\label{automorphismP}
\end{cases}
\end{equation}
Notice how the transition cocycle $\al: U\arr \R$ acts on the left as in the general case, being in this case left and right trivial since the group is commutative.
The canonical (right) action is denoted by $R_a: P\arr P$ and it acts locally as $R_a: P\arr P: [x, l]\mapsto [x, l+a]$.

Since transition functions (\ref{automorphismP}) are affine transformations, the $\R$-principal bundle $P$ is also an affine bundle. 
As an affine bundle it allows global sections, which make it necessarily trivial as a principal bundle.
Then it is not restrictive to consider $P$ trivial, i.e.~taking $P=M\times \R$.

An automorphism of $P$ is a pair of maps $(\vp, \Phi)$ acting as
\begin{equation}
\begindc{\commdiag}[1]
\obj(110,70)[P1]{$P$}
\obj(170,70)[P2]{$P$}
\obj(110,30)[M1]{$M$}
\obj(170,30)[M2]{$M$}
\mor{P1}{M1}{$\pi$}
\mor{P2}{M2}{$\pi$}
\mor{P1}{P2}{$\Phi$}
\mor{M1}{M2}{$\vp$}
\enddc
\end{equation}
and commuting with the (right) action, i.e.~$\Phi\circ R_a= R_a\circ \Phi$. Locally, an automorphism of $P$ is then in the form
\begin{equation}
\begin{cases}
x'^\mu= \vp^\mu(x)\\
l'= \om(x) +l
\end{cases}
\label{AutTransf}
\end{equation}
for some local pointwise element of the group  $\om(x)\in \R$.

Fibered coordinates define a basis $(\del_\mu, \del)$ of tangent vectors to $P$. In this case the vector $\rho=\del$ is a (right) invariant pointwise basis for vertical vectors.
Accordingly, an infinitesimal generator of automorphisms on $P$ is a vector field in the form
\begin{equation}
\Xi=\xi^\mu(x)\del_\mu + \ze(x) \rho \label{vectorfield}
\end{equation}
which projects onto the spacetime vector field $\xi=\xi^\mu(x)\del_\mu$.

%Let us fix a signature $\eta=(r, s)$  for manifolds of dimension $m=r+s$.
Let us denote by $B(\eta)$ the space of all symmetric non-degenerate bilinear forms  of signature $\eta$; the set $B(\eta)$ is an open set in the vector space of symmetric $2$-tensors $S_2(\R^m)\simeq \R^{\Frac[m(m+1)/2]}$, parametrized by coordinates $g_{ab}$.
We can define a left action of the group $\R\times \GL(m)$ on $B(\eta)$ by
\begin{equation}
\la:\R\times \GL(m)\times B(\eta)\arr B(\eta): (\om, J_a^b, g_{ab})\mapsto g'_{ab}= e^\om \> \bar J_a^c g_{cd} \bar J^d_b
\end{equation}
Let us stress that this action preserves non-degeneracy and signature so that it is really a group action on $B(\eta)$.

We can build the associated bundle
\begin{equation}
C:=P\times L(M)\times_\la B(\eta)
\end{equation}
to be used as configuration bundle. 
Here we denoted the general frame bundle by $L(M)$.
Local coordinates on $C$ are in the form $(x^\mu, g_{\mu\nu})$ and they transform under automorphisms (\ref{AutTransf}) as
\begin{equation}
\begin{cases}
x'^\mu= \vp^\mu(x)\\
g'_{\mu\nu} = e^\om \bar J_\mu^\al g_{\al\be} \bar J^\be_\nu
\end{cases}
\end{equation}
This embeds the group $\Aut(P)$ into the group $\Aut(C)$ and the subgroup $\Aut(P)\subset \Aut(C)$ will be below required to act by symmetries of the system.
Notice that while $\Aut_V(P)\subset \Aut(C)$ forms a subgroup of transformations, called {\it vertical automorphisms} or {\it proper gauge transformations},
spacetime diffeomorphisms $\Diff(M)$ are not embedded into $\Aut(C)$ in general. On the contrary the group $\Aut(P)\subset \Aut(C)$ {\it projects} onto $\Diff(M)$.
Accordingly, one cannot say that spacetime diffeomorphisms act on fields, even though transformations in $\Aut(P)$, also called {\it generalized gauge transformations},
take into account both spacetime diffeomorphisms and proper gauge transformations.
 
The infinitesimal transformations of fields under generalized gauge transformations are encoded by Lie derivatives, namely
\begin{equation}
\Lie_\Xi g_{\mu\nu} = \xi^\al D_\al g_{\mu\nu} + \na^{(\Ga)}_\mu \xi^\al g_{\al\nu}+ \na^{(\Ga)}_\nu \xi^\al g_{\mu\al} + \ze_V g_{\mu\nu}
\label{Lie}
\end{equation}
where we select a connection $\Ga$ on the spacetime $M$ and a principal connection $\te$ on $P$, where $\na^{(\Ga)}_\mu \xi^\al= \del_\mu \xi^\al + \Ga^\al_{\la\mu} \xi^\la$
is the usual covariant derivative of $\xi$ with respect to the spacetime connection $\Ga$, where we set $\ze_V= \ze + \xi^\mu \te_\mu$ for the vertical part of $\Xi$
and where we defined the gauge covariant derivative (with respect to $\Ga$ and $\te$) of fields as
\begin{equation}
D_\al g_{\mu\nu} = \del_\al g_{\mu\nu} - \Ga^\la_{\mu\al} g_{\la\nu} - \Ga^\la_{\nu\al} g_{\mu\la} - \te_\al g_{\mu\nu}
\end{equation}
Notice how the Lie derivative $\Lie_\Xi g_{\mu\nu}$ does not depend on connections while the single terms in it do.

Although the configuration bundle $C$ has coordinates $(x^\mu, g_{\mu\nu})$ it does not have to be confused with the bundle $\Met(M; \eta)$ of metrics of signature 
$\eta$ on $M$.  Global sections of $C$ %not  in fact global metrics on $M$, rather they 
are a family of metrics %defined on an open covering of $M$ which differ on the overlaps by a 
modulo conformal transformations.
%The set of global sections of $C$ is in fact richer than the set of global metrics (of signature $\eta$). 
Let us stress that in this context conformal transformations will act as
gauge transformations and thence they do not affect the physical content of fields.
One could say that a global section of $C$ is an implementation of a conformal structure on $M$. For example, a section of $C$ allows to define all the geometrical structures (light cones, spacelike, timelike, lightlike directions) which are conformally invariant. On the other hand it does not define the length of vectors (which instead depends on a conformal representative).

%%%%%%%%%%%%%%%%%%%%%%%%%%%%%%%%%%%%%%
\section{Conformal gravity}

In a field theory based on the kinematics described above, one chooses a Lagrangian to provide a dynamics.
The Lagrangian is required to depend on $g_{\mu\nu}$ together with derivatives up to some finite order $k$ ($k=2$ for our convenience here).
The dynamics is required to be gauge covariant, i.e.~transformations in $Aut(P)$ are required to be Lagrangian symmetries.
Such a field theory on $C$ is a gauge natural theory (see \cite{FatibeneLibro}); accordingly, the theory automatically allows a superpotential and an associated conserved quantity obtained by integration on an $(m-2)$-surface in spacetime {\it \'a la} Gauss.

We need to find out a dynamics which is invariant with respect to conformal transformations. One can consider a Lagrangian in the form:
\begin{align}
L^\ast&=\sqrt{g} \left(a\, R^2 + b\, R_{\mu\nu}R^{\mu\nu} + c \, R^\al{}_{\la\mu\nu} R_\al{}^{\la\mu\nu} \right) d \si \notag \\
&= \left(a L_1+b L_2+c L_3 \right) d\si
\label{totalL}
\end{align}
where $\sqrt{g}$ denotes the square root of the absolute value of the determinant of $g_{\mu\nu}$ and $d\si$ is the local basis for $m$-forms on $M$ induced by coordinates. Indices are raised and lowered by the field $g_{\mu\nu}$, we set $R_{\be\mu}:=R^\al{}_{\be\al\mu}$ for the Ricci tensor and $R:= g^{\be\mu} R_{\be\mu}$ for the scalar curvature. The coefficients $a, b$ and $c$ are real and they have to be determined so that the Lagrangian \eqref{totalL} turns out to be conformally invariant. 

Let us hereafter restrict to the case $m=4$ and Lorentzian signature $\eta=(3,1)$.
By a standard Utiyama technique (see \cite{FatibeneLibro}, \cite{Uti},  \cite{Jan}) one can directly show that the only Lagrangian
conformally invariant is 
\beq
L=3a \sqrt{g} \left( \frac{1}{3} R^2- 2 R_{\mu\nu}R^{\mu\nu}+R_{\al\be\mu\nu}R^{\al\be\mu\nu} \right) d\si 
\label{totalLconf}
\eeq
One can observe that the quantity in brackets is just:
\beq
W_{\al\be\mu\nu}W^{\al\be\mu\nu}= R_{\al\be\mu\nu} R^{\al\be\mu\nu} - 2 R_{\al\be} R^{\al\be} + \Frac[1/3] R^2
\eeq
where $W_{\al\be\mu\nu}$ is the well-known Weyl tensor defined by
\begin{equation}
W_{\al\be\mu\nu} = R_{\al\be\mu\nu}  - \(g_{\al[\mu} R_{\nu]\be} - g_{\be[\mu} R_{\nu]\al}\) +\Frac[1/3]R g_{\al[\mu} g_{\nu]\be}
\label{WeylTensor}
\end{equation}
The Lagrangian \eqref{totalLconf} is often considered in the literature in the reduced form
\begin{equation}
\tilde L= 2 R_{\mu\nu}R^{\mu\nu} - \frac{2}{3} R^2
 \label{recastedL}
\end{equation}
which is obtained from the general form (\ref{totalLconf}) by discarding a  Gauss-Bonnet (GB) term $G=R_{\la\mu\nu k}R^{\la\mu\nu k} - 4R_{\mu\nu}R^{\mu\nu} + R^2$. 
Discarding a GB term has no effect on field equations (since GB is known to be a total divergence) while it effects covariance.
Indeed, neither the GB term nor the remaining Lagrangian $\tilde L$ are separately conformally invariant.
Since we are here interested especially in conservation laws, subtracting GB contributions is a particularly bad idea.
For that reason we keep Lagrangian in the form (\ref{totalLconf}).
We leave for a late comment the issue of conservation laws for Lagrangian (\ref{recastedL}) which can be easily obtained by the general framework in the end; see Section \ref{sec:komar}.

The variation of the partial Lagrangians $L_i$ (as defined in equation (\ref{totalL}))
can be canonically split by the first variation formula  as
\beq
\begin{aligned}
\de L_1=&\sqrt{g}\(2RR_{\mu\nu} -\frac{1}{2}g_{\mu\nu}R^2+2\square R g_{\mu\nu} -2 \na_{(\mu}\na_{\nu)}R\)\de g^{\mu\nu}\\
+& \sqrt{g}d\Big[
	\Big(2\na_{\eta}R \left( g^{\eta\la}g^{\mu\nu} - g^{\eta(\nu} g^{\mu)\la} \right) \de g_{\mu\nu}\\
 	&\qquad+2R \left( g^{\rho(\mu}g^{\nu)\la} - g^{\la\rho}g^{\mu\nu} \right) \na_\rho \de g_{\mu\nu}\Big) \otimes d\si_\la \Big]
\label{fieldeq1}
\end{aligned}
\eeq
\beq
\begin{aligned}
\de L_2=&\sqrt{g}\( 2R_{\mu\al} R_\nu {}^\al - \frac{1}{2} S g_{\mu\nu} - 2\na_\la \na_{(\mu} R_{\nu)} {}^\la+ \square R_{\mu\nu} 
+\frac{1}{2} \square R g_{\mu\nu}\) \de g^{\mu\nu}\\
+& \sqrt{g}d\Big[
	\Big(\( -2\na_{(\mu}R^\la{}_{\nu)} + \na^\la R_{\mu\nu} + \hbox{$\frac{1}{2}$} \na^\la R g_{\mu\nu} \right) \de g_{\mu\nu}\\
 	&\qquad+\(2R^{\rho\nu}g^{\la\mu} - R^{\mu\nu} g^{\la\rho} - R^{\la\rho}g^{\mu\nu} \right) \na_\rho \de g_{\mu\nu}\Big) \otimes d\si_\la \Big]
\end{aligned}
\eeq
\beq
\begin{aligned}
\de L_3=\sqrt{g}&\( -\hbox{$\frac{1}{2}$} K g_{\mu\nu} + 2 R_{\mu\al\be\ga}R_\nu{}^{\al\be\ga} - 4 \na_\la \na_\ep R^\la{}_{\mu\nu}{}^\ep \) \de g^{\mu\nu}\\
+\sqrt{g}& d\Big[
	\Big(-4\na_\ep R^{\la\mu\nu\ep}   \de g_{\mu\nu} +4R^{\rho(\mu\nu)\la}  \na_\rho \de g_{\mu\nu}\Big) \otimes d\si_\la \Big]
\end{aligned}
\eeq
where we set $S:=R^{\al\be} R_{\al\be}$ and $K:=R^{\al\be\ga\de}R_{\al\be\ga\de}$.

The volume parts contribute to field equations, while the boundary parts  enter in conservation laws (see Section \ref{sec:komar}).
Field equations for the Lagrangian \eqref{totalLconf} are then:
\begin{multline}
E_{\mu\nu}:=a \sqrt{g} \left( 2RR_{\mu\nu} -\frac{1}{2}g_{\mu\nu}R^2+2\square R g_{\mu\nu} -2 \na_{(\mu}\na_{\nu)}R \right) \\
- 6 a \sqrt{g}\left( R_{\mu\al} R_\nu {}^\al - \frac{1}{4} S g_{\mu\nu} - \na_\la \na_{(\mu} R_{\nu)} {}^\la+\frac{1}{2} \square R_{\mu\nu} +\frac{1}{4} \square R g_{\mu\nu} \right) \\
+3 a \sqrt{g} \left( -\frac{1}{2} K g_{\mu\nu} + 2 R_{\mu\al\be\ga}R_\nu{}^{\al\be\ga} - 4 \na_\la \na_\ep R^\la{}_{\mu\nu}{}^\ep \right) =0
\end{multline}
Since $E_{\mu\nu}$ is obtained from a Lagrangian that is both generally covariant with respect to changes of coordinates and conformally invariant, it is kinematically covariantly conserved and traceless and obeys $\na^\mu E_{\mu\nu}=0$, $E_{\mu\nu}g^{\mu\nu}=0$. Thus, from the latter, we have another constraint on the constants $a,b,c$ of the total Lagrangian \eqref{totalL}:
\beq
3a +b+c=0
\label{ConfEqInvariance}
\eeq
Both the Lagrangian \eqref{totalLconf} and Gauss-Bonnet (as well as of course the Lagrangian density $ \tilde L= L-\sqrt{g} G= \sqrt{g}\(2S-\frac{2}{3} R^2 \)$
which is used in \cite{Mannheim:1989}) verify this condition. However, this condition is more general than the condition for conformally invariant Lagrangians.
In fact the Lagrangian (\ref{totalLconf}) is the only dynamics which is conformally invariant, while condition
(\ref{ConfEqInvariance}) identifies dynamics for which conformal transformations are symmetries of the equations (i.e.~generalized Lagrangian symmetries).

%As far as solutions of field equations are concerned, all vacuum solutions of Einstein theory have $R_{\mu\nu}=0$ and are thence solution of conformal gravity.

Of course, when a metric $g_{\mu\nu}$ is a solution of conformal gravity, then all the metrics which are conformal to it, namely, all 
$\tilde g_{\mu\nu}= e^{\om(x)}\cdot g_{\mu\nu}$, are solutions as well. This is a consequence of the fact that conformal transformations are gauge symmetries.

The solutions of conformal gravity which are stationary and spherically symmetric (up to a generic conformal factor) are \cite{Mannheim:1989}
\begin{equation}
g = \Phi(t, r, \te, \phi)\(-A(r) dt^2 + \Frac[1/A(r)] dr^2 + r^2 (d \te^2 + \sin^2(\te) 
d\phi^2)\)
\label{solutionCG}
\end{equation}
where we set
\begin{equation}
A(r)= 1- \frac{\be(2-3\be\ga)}{r} -3\be\ga+\ga r - k r^2 \label{symsolution}
\end{equation}

In general, there are allowed and forbidden regions for the coordinate $r$ depending on the value of the parameters $(\be, \ga, k)$.
Generally, one has that for $k>0$, $r$ cannot go to $+\infty$; while for $k<0$,  it can.

Let us remark that the physical meaning of constants appearing in this solution needs to be clarified. Mathematically they appear as integration constants
while they appear in the solution as physically motivated constants in other contexts. For example the constant $k$ appears as a cosmological constant in Schwarzschild-de-Sitter solutions though here it appears in the solution as an integration constant (i.e.~in principle one has solutions with any value of it) while in standard GR it appears in the Lagrangian and hence has a definite value imposed at the level of dynamics.
Similarly, it is not clear which combination of constants plays the role of the physical mass (of the point mass at the origin) which would be essential in analyzing applications to gravitational lensing.

%%%%%%%%%%% CORRENTI
\section{Superpotential}
\label{sec:komar}

If one considers a generator of a pure conformal transformation, namely $\Xi=\ze(x)\rho$, the Lie derivative of the metric reads as
\begin{equation}
\Lie_\Xi g^{\mu\nu} = \ze g^{\mu\nu}
\end{equation}
which depends on $\ze$ though not on its derivatives.

\begin{comment}
%LIE DERIVATIVE
\begin{align}
\Lie_\Xi \Ga^\la_{\mu\nu}=&\frac{1}{2} \left( g^{\la\be}g_{\mu\nu} -2 \de^\la_{(\mu} \de^\be_{\nu)} \right) \na_\be \ze \label{Liegamma},\\
\Lie_\Xi u^\la_{\mu\nu}=& \frac{1}{2} \left( g^{\la\be}g_{\mu\nu} + 2 \de^\la_{(\mu} \de^\be_{\nu)} \right) \na_\be \ze \label{Lieu}.
%\Lie_\Xi R_{\mu\nu}=&\frac{1}{2} g_{\mu\nu} \Box \ze + \na_{\mu\nu} \ze.
\end{align}
\end{comment}
%
%
\noindent Accordingly, following the general theory (see \cite{FatibeneLibro}) for the conformal Lagrangian (\ref{totalLconf}) we have the Noether current:
\begin{equation}
\Noe_{\textup{conf}}= \sqrt{g}\(T^\la \ze + T^{\la\ep} \na_\ep \ze\) d\si_\la
\end{equation}
where:
\begin{align}
T^\la=F^{\la\mu\nu}g_{\mu\nu}=&\,0 ,\label{nullreduced}\\
T^{\la\ep}=F^{\la\ep\mu\nu}g_{\mu\nu}= &\,a\left( R g^{\mu\nu} - 6 R^{\mu\nu} \right) \left( g^{\la\ep}g_{\mu\nu} + 2 \de^\ep_\mu \de^\la_\nu \right)+ \notag \\
&+6 a R_\al{}^{\be\la\nu} \left( g^{\al\ep}g_{\be\nu} - \de^\al_\nu \de^\ep_\be \right) =0 \label{nullKomar}
\end{align}
and 
\begin{align}
F^{\la\mu\nu}&=2a\na_{\eta}R \( g^{\eta\la}g^{\mu\nu} - g^{\eta(\nu} g^{\mu)\la} \)+\\
&-6a\(-2\na_{(\mu}R^\la{}_{\nu)} + \na^\la R_{\mu\nu} + \hbox{$\frac{1}{2}$} \na^\la R g_{\mu\nu}\) 
-12a\na_\ep R^{\la\mu\nu\ep}  \\ 
F^{\la\rho\mu\nu}&= 2aR \( g^{\rho(\mu}g^{\nu)\la} - g^{\la\rho}g^{\mu\nu} \right)
-6a\(2R^{\rho\nu}g^{\la\mu} - R^{\mu\nu} g^{\la\rho} - R^{\la\rho}g^{\mu\nu} \right) +\\
&+ 12aR^{\rho(\mu\nu)\la}
\end{align}

\noindent The Noether current identically vanishes and it
can be trivially split as $\Noe=\tilde{\Noe} + \textup{d}\mathcal{U}$. Then, equations \eqref{nullreduced}-\eqref{nullKomar} mean that both the reduced current $\tilde{\Noe}$, which in general vanishes on-shell, and the superpotential $\UU$ are in fact identically zero off-shell. The fact that the superpotential relative to the conformal symmetry is zero means that conformal symmetry gives null `conserved charge'.

\begin{comment}
The vanishing Noether charge for conformal symmetry is not unexpected. Moreover, if we introduce the \emph{work form} \cite{FatibeneLibro}:
\beq
W:=E_{\mu\nu} \Lie g^{\mu\nu}
\eeq
using the equation \eqref{Liegmunu} and the fact that field equations are traceless, we have:
\beq
W=E_{\mu\nu}g^{\mu\nu} \ze=0
\eeq
Since:
\beq
\textup{div} \Noe=W=0, \quad \Noe=\left( T^\la \ze + T^{\la\al}\na_\al \ze \right) d\si_\la,
\eeq
we have:
\beq
\na_\la T^\la \ze + \left( T^\la +\na_\al T^{\al\la}\right) \na_\la \ze + T^{\la\al}\na_{\al\la} \ze=0.
\eeq
\end{comment}

%corrente diff
If one considers the contribution of an infinitesimal generator of diffeomorphisms, $\xi=\xi^\mu(x) \del_\mu$, the Noether current again for the Lagrangian (\ref{totalLconf}) is:
\beq
\Noe_{\textup{diff}}=\sqrt{g}\left( T^\la{}_\ep \xi^\ep + T^{\la\mu}{}_\ep \na_\mu \xi^\ep + T^{\la\mu\nu}{}_\ep \na_{\mu\nu} \xi^\ep \right) d\si_\la
\eeq
where
\begin{align}
T^\la{}_\ep=& \,2a\left(R g^{\mu\nu} - 6R^{\mu\nu}\right) \left( -R^\la{}_{\mu\nu\ep}+ \de^\la_\mu R_{\nu\ep} \right)-12a R_\mu{}^{\be\la\nu}R^\mu{}_{(\be\nu)\ep}+ \\
&+a\left( R^2-6S+3K\right)\de^\la_\ep \notag \\
T^{\la\mu}{}_\ep=&-2a \na^\mu R\de^\la_\ep -2 a\na_\ep R g^{\la\mu}-2a \na^\la R \de^\mu_\ep+ 12a \na^\la R^\mu {}_\ep \\
T^{\la\mu\nu}{}_\ep=& \, 2aRg^{\mu\nu}\de^\la_\ep -2aR g^{\la(\mu}\de^{\nu)}_\ep - 12a R^{\mu\nu} \de^\la_\ep + 12 aR^{\la(\mu}\de^{\nu)}_\ep +\notag \\
&- 12a R_\ep{}^{(\mu\nu)\la}
\end{align}
The superpotential is (see \cite{FatibeneLibro}):
\beq
\UU^{\la\mu}=\frac{1}{2} \left\{ \left( T^{[\la\mu]}{}_\ep -\frac{2}{3} \na_\nu T^{[\la\mu]\nu}{}_\ep \right)\xi^\ep+ \frac{4}{3} T^{[\la\mu]\nu}{}_\ep \na_\nu \xi^\ep \right\}
\eeq
In order to compute it we need the following objects:
\begin{align}
T^{(\la\mu\nu)}{}_\ep&=0 \\
T^{[\la\mu]}{}_\ep&= 12a \na^{[\la} R^{\mu]}{}_\ep \\
T^{[\la\mu]\nu}{}_\ep&= 3 aR g^{\nu[\mu} \de^{\la]}_\ep + 18aR^{\nu[\la} \de^{\mu]}_\ep + 6a R_\ep{}^{[\mu\la]\nu} + 6a R_\ep{}^{\nu\la\mu} \\
\na_\nu T^{[\la\mu]\nu}{}_\ep&= 6a \na^{[\la} R \de^{\mu]}_\ep - 18a \na^{[\la}R^{\mu]}{}_\ep
\end{align}
Then $\UU$ is:
\begin{multline}
\UU=\sqrt{g}\UU^{\la\mu}d \si_{\la\mu}=a\sqrt{g}\left\{ \left( 12 \na^{[\la} R^{\mu]}{}_\ep - 2\na^{[\la} R \de^{\mu]}_\ep \right) \xi^\ep + \right. \\
+ \left. 2 \left( R g^{\nu[\mu} \de^{\la]}_\ep + 6 R^{\nu[\la} \de^{\mu]}_\ep + 3 R_\ep{}^{\nu\la\mu}\right) \na_\nu \xi^\ep \right\} d \si_{\la\mu} \label{Komar}
\end{multline}

The superpotential is conformally invariant, meaning it is associated to any conformal metric $\tilde g=\Phi(r) \cdot g$.

More generally, we can compute the conservation laws associated to the infinitesimal generator $\xi$ of spacetime diffeomorphisms for Lagrangian (\ref{totalL})
which is in fact generally covariant even when it is not conformally invariant.

For the superpotential of the Lagrangian (\ref{totalL}) with respect to the generator $\xi$, one has 
\begin{multline}
\UU^\ast=\sqrt{g}(\UU^\ast)^{\la\mu}d \si_{\la\mu}=\sqrt{g}\left\{ \left( 2(b+4c) \na^{[\la} R^{\mu]}{}_\ep +(4a+b)\na^{[\la} R \de^{\mu]}_\ep \right) \xi^\ep + \right. \\
+ \left. 2 \left( aR g^{\nu[\mu} \de^{\la]}_\ep -b R^{\nu[\la} \de^{\mu]}_\ep + c R_\ep{}^{\nu\la\mu}\right) \na_\nu \xi^\ep \right\} d \si_{\la\mu}
\end{multline}
which in fact specializes to the superpotential (\ref{Komar}) by setting $b=-6a$ and $c=3a$ as to obtain the Lagrangian (\ref{totalLconf}).

By specializing instead to the Lagrangian $\tilde L$ given by (\ref{recastedL}) one obtains
\begin{multline}
\tilde \UU=\sqrt{g}\tilde \UU^{\la\mu}d \si_{\la\mu}=\sqrt{g}\left\{ \left( 4 \na^{[\la} R^{\mu]}{}_\ep -\frac{2}{3}\na^{[\la} R \de^{\mu]}_\ep \right) \xi^\ep + \right. \\
- \left. 4 \left( \frac{1}{3}R g^{\nu[\mu} \de^{\la]}_\ep + R^{\nu[\la} \de^{\mu]}_\ep \right) \na_\nu \xi^\ep \right\} d \si_{\la\mu}\label{ReducedKomar}
\end{multline}

Notice how different Lagrangians give different superpotentials even when they are dynamically equivalent as it happens for Lagrangians 
  $L$ given by (\ref{totalLconf}) and $\tilde L$ given by (\ref{recastedL}); see \cite{Mann}\cite{FFF_Energy}.

 %%%%%%%%%%%%%%%%%%%%%%%%%%
\section{Conservation laws}

Conserved charges are obtained from conservation laws by integrating the superpotential  on  surfaces, e.g.~at $t=const$ and $r=const$.
We shall hereafter consider the case $k<0$ and so that we are allowed to let the radius of the sphere tend to infinity.

Let us fix $\xi= \del_0$;  the only non-vanishing components of the superpotential, for the spherically symmetric solution in \eqref{solutionCG}, are:
\begin{multline}
\UU^{tr}=- \UU^{rt}=a\frac{6 \be\sin(\te)}{r^3} \left( \be(2-3\be\ga)^2 + 3\be\ga (2  -3  \be\ga)r  + 3\be\ga^2 r^2 + \right. \\
\left.     +(6k \be\ga-\ga^2  -4 k )r^3 \right) 
\label{Komartr}
\end{multline}

We have:
\beq
Q = \int_{S^2} \!\!  \, \UU^{\la\mu}d\si_{\la\mu}=\frac{1}{4\pi} \int_0^{2\pi}\!\! \kern-6pt d\varphi \! \int_0^\pi \!\! d\te \, \UU^{tr}
\label{Cintegral}
\eeq
%{\bf (manca una normalizzazione $1/n\pi$ vd maple)}
Then one obtains:
\begin{multline}
Q=-6 \be \left( 4k + \ga^2 - 6k \be \ga \right) + \frac{18 \be^2\ga^2}{r} + \frac{18\be^2 \ga (2-3\be\ga)}{r^2} +\\
+ \frac{6 \be^2 (2- 3 \be \ga)^2}{r^3}
\end{multline}

This quantity is conformally invariant. If one lets $r$ tend to infinity, then:
\beq
Q_{\infty} = -6\be \left(4k+\ga^2-6k\be\ga \right)
\label{Noecharge}
\eeq

Also these quantities are conformally invariant, hence solutions with different values of $Q_{\infty}$ cannot be conformally equivalent.
For example, for $\be=0$ all solutions have $Q_{\infty}=0$.  In fact one can show that the solutions corresponding to $\be=0$, i.e.~with
\beq
A_0(r)=1+\ga r - k r^2
\label{ConfFlat} 
\eeq
are in fact conformally flat since their Weyl tensor vanishes.

On the contrary when $\be\not=0$ different values of $\ga$ provide different values of $Q_{\infty}$ and hence they are not conformally equivalent.

The same metric is a solution for field equations of the Lagrangian $\tilde L$ given by (\ref{recastedL}) as well.
The integral of the superpotential $\tilde \UU$ given by (\ref{ReducedKomar}) gives 
\begin{multline}
\tilde Q=4k^2r^3-6k\gamma r^2+2\gamma(\gamma+6k\beta)r + 2\beta( -4\gamma^2+k( 3\beta\gamma-2))+6\frac{\beta^2\gamma^2}{r}
\end{multline}
Accordingly, conserved quantities of the same solution for different equivalent Lagrangians do in fact differ.

Notice that $\tilde Q$ is not bounded as $r\arr \infty$. The GB term which is added to $\tilde L$ to obtain $L$ acts in fact as a regularizing boundary term; 
see \cite{FatibeneAugmented2005}.

\section{Conclusion and Perspectives}

We have studied  conformal theories of gravity in the framework of gauge natural theories that have a general method for obtaining conserved quantities. We have considered a Lagrangian \eqref{totalL} which is a linear combination of three possible quadratic scalars containing the fields (the metric) and their derivatives up to order 2. We have shown that the Lagrangian \eqref{totalL} has a unique choice  (up to a global factor) of the constants $a,b$ and $c$ that makes it conformally invariant.
%And we have shown that this choice is, up to a constant, the Weyl Lagrangian.

We have derived its field equations and conserved currents.
The conservation laws and superpotentials are interesting on their own.
Moreover, they could give information on the physical meaning of the integration constants in the static spherically symmetric solution \eqref{solutionCG}. We have pointed out that the conserved charge associated to spacetime diffeomorphisms is a conformally invariant quantity. Thus, its physical meaning is not trivial since the standard operative definition of mass is not.
% up to a constant

Conserved quantities need also a careful analysis of asymptotic structure of the solutions.
In general one also needs a number of assumptions about the family parameters in order to control the asymptotics and make it independent of the family parameters.
Here the situation is even more difficult to control than in standard metric theories; here one also has a conformal gauge invariance so that probably one can control asymptotics modulo gauge transformations.
In fact one can directly show that in the case of $k<0$ two elements of the family have the same asymptotics at $r=r_M$ only if they coincide everywhere.
Conformal gauge could provide the freedom for non-trivial solutions and allow to compute relative quantities;
 \cite{FatibeneAugmented2005}).
 
Although this needs a non-trivial generalization of the theory it may give hints about the meaning of the parameters.
Only after having a better understanding of  the relations among the three constants we can deal with light geodesics and consequential deflection angle and lensing. Light bending and deflection angles in conformal gravity are studied in \cite{Edery:1997} and \cite{Pireaux:2004}. 
However,  we think that if we cannot reliably relate $\be,\ga$ and $k$ to any physical quantities, for example, we cannot perturbatively solve geodesics equation or know which solution corresponds to the presence of a massive body or to the background (turning off the ``mass'');
see  \cite{SultanaKazanas} and \cite{CattaniScalia2013}.

Future investigation needs to be devoted to discuss the relation between the gauge natural approach and the more traditional natural approach in which conformal transformations are treated as special diffeomorphisms which can be lifted to fields.

\section*{Acknowledgments}
We wish to thank A.Diaferio, N.Fornengo and C.Germani for useful discussions and comments. We also acknowledge the contribution of INFN (Iniziativa Specifica QGSKY), the local research project {\it  Metodi Geometrici in Fisica Matematica e Applicazioni (2013)} of Dipartimento di Matematica of University of Torino (Italy). This paper is also supported by INdAM-GNFM. One of us (M.C.) acknowledges partial support from the INFN grant InDark and from the grant Progetti di Ateneo/CSP TO\textunderscore Call2\textunderscore 2012\textunderscore 011 `Marco Polo' of the University of Torino and  thanks M. Meineri for valuable suggestions and support.

%thank Marco

\printbibliography

\end{document}